# Dual equilibrium in a finite aspect ratio tokamak

P.-A. Gourdain[1], S. C. Cowley[1] [2], J.-N. Leboeuf[3]

[1] Department of Physics and Astronomy, University of California Los Angeles, CA 90095, USA

[2] Department of Physics, Imperial College, Prince Consort Road, London SW7 2AZ, UK

[3] JNL Scientific, Casa Grande, AZ 85294, USA

A new approach to high pressure magnetically-confined plasmas is necessary to design efficient fusion devices. This paper presents an equilibrium combining two solutions of the Grad-Shafranov equation, which describes the magnetohydrodynamic equilibrium in toroidal geometry. The outer equilibrium is paramagnetic and confines the inner equilibrium, whose strong diamagnetism permits to balance large pressure gradients. The existence of both equilibria in the same volume yields a dual equilibrium structure. Their combination also improves free-boundary mode stability.

The most promising candidate to a large-scale fusion reactor is the tokamak concept, where a closed magnetic topology confines a hot ionized gas, called plasma, where electrons and ions are not bound together due to energetic collisions. To reduce particle loss, a strong toroidal magnetic field $B_\phi$ ($\phi$ denotes the toroidal axisymmetric direction) is used and effectively locks both charged species on magnetic field lines. This results in relative thermal insulation. However turbulence and collisions between particles degrade confinement. Whilst the plasma core is hot, the edge remains relatively cold and a pressure gradient exists across the plasma section. In order to obtain a magnetohydrodynamic (MHD) equilibrium, a toroidal current density $J_\phi$, runs inside the plasma and generates a poloidal field $B_P$. The Lorentz force gives rise to the inward radial force necessary to balance the pressure gradient. Paradoxically the toroidal field $B_\phi$ does not play any role in this balancing act. However it limits the maximum value of $J_\phi$, in turn controlling the maximum allowable pressure. As a consequence, the fusion power follows the scaling law given in Eq. (1),

$$P_{fusion} \propto \langle \beta \rangle^2 B^4 a^3 A. \qquad (1)$$

$a$ is the plasma minor radius, $R$ is the plasma major radius and $A$ is the aspect ratio given by $R/a$. $B$ is the total field inside the plasma and $\beta$ measures the efficiency of kinetic pressure confinement by magnetic fields, i.e.

$$\beta = 2\mu_0 \frac{p}{B^2} \text{ and } \langle \beta \rangle = 2\mu_0 \left\langle \frac{p}{B^2} \right\rangle. \qquad (2)$$

<.> denotes the volume average quantity and $p$ the plasma kinetic pressure. To obtain an attractive fusion reactor design, Eq. (1) shows that $\beta$ has to be maximized, while reactor costs tend to reduce $B_\phi$ since its creation is expensive. Unfortunately the Troyon limit [1] restricts the allowable plasma $\langle\beta\rangle$ to a few percent. Beyond a critical value $\beta_c$, MHD disturbances, or modes, perturb the axisymmetry of the plasma, leading to loss in confinement and, ultimately, plasma disruptions. The normal $\beta$, defined by

$$\beta_N = \frac{\langle\beta\rangle(\%) a(m) B_\phi(T)}{I_p(MA)}, \qquad (3)$$

is a relative measurement of plasma stability. Here, $I_P$ is the total toroidal plasma current. Instabilities typically occur for $\beta_N$ above 2.5 or 3. This requirement is found to be quite robust in any experiment running with conventional current profiles. However reactor economics requires pressures larger than presently achievable in conventional tokamaks. Previous research has demonstrated that high pressure equilibria exist and are stable to fixed boundary modes $n$ = 1, 2 and 3 [2], internal asymmetries which lead to plasma disruptions. Unfortunately free-boundary modes, which are external asymmetries, remain unstable. Their stabilization would require a perfectly conducting vacuum vessel wall next to the plasma edge, a solution which is not realistic. This paper proposes to identify the cause of this external instability and presents a possible remedy.

The extended energy principle [3] assesses the nature of free-boundary mode stability by studying the perturbed plasma and vacuum energies caused by infinitesimal displacements. These displacements generate a total perturbed energy $\delta W_{Total}$, which is a volume integral over the whole plasma-vacuum system. Its mathematical nature enables integration by part, effectively splitting the system into two volume integrals: over the plasma, yielding the perturbed energy $\delta W_{Plasma}$, and over the whole vacuum region, yielding $\delta W_{Vacuum}$. We will suppose here that no currents run on the plasma edge. The system is stable if and only if the total perturbed energy is positive for *any* infinitesimal displacement. For an displacements $\xi_\perp$ locally perpendicular to the magnetic field, we can express the perturbed energy using in the following form [4]

$$\delta W_{Plasma} = \frac{1}{2} \int_{Plasma} \left[ \frac{\mathbf{Q}_\perp^2}{\mu_0} + \frac{B^2}{\mu_0} |\nabla \cdot \xi_\perp + 2\xi_\perp \cdot \kappa|^2 \right.$$
$$\left. -2[\xi_\perp \cdot \nabla p][\kappa \cdot \xi_\perp] - J_\parallel [\xi_\perp \times \mathbf{b}] \cdot \mathbf{Q}_\perp \right] d\tau \quad (4)$$

$\mathbf{Q}_\perp = \nabla \times (\xi_\perp \times \mathbf{B})$ is the perturbed plasma magnetic field, **b** correspond to the magnetic field direction and the curvature of the magnetic field lines is given by $\kappa = \mathbf{b}.\nabla \mathbf{b}$. While it is evident from Eq (4) that large pressures will lead to negative perturbed energies, finite aspect ratio tokamaks also suffer from a handicapping side effect at high pressure, diamagnetism [ 5 ]. The loss in magnetic field compressibility ($2^d$ term in Eq. (4)) is an inconvenient by-product, which further reduces the perturbed plasma energy. Stability is sensitive to this term since the magnetic field strength has a quadratic contribution. Indeed purely diamagnetic plasmas are free boundary unstable [2]. We can circumvent this conundrum since Eq. (4) is also an integral. Thus this equation can be split again into two parts. If we allow diamagnetism in the core region of the plasma, to maximize pressure, and if we impose paramagnetism in the outer region, to maximize magnetic field, then we should be able to increase the total perturbed plasma energy by tailoring the equilibrium in both regions. As we show later on, this approach should yield free-boundary mode stability. The next task is to investigate if it is possible to also split the MHD equilibrium into a diamagnetic and paramagnetic equilibrium, as we did for the volume integral. The axisymmetric MHD equilibrium of a plasma is ruled by the Grad-Shafranov (GSh) equation [6]

$$\mu_0 J_\phi(R,Z) = -\left( \mu_0 R \frac{dp}{d\psi} + \frac{1}{2R} \frac{dF^2}{d\psi} \right) \quad (5)$$

given here in the (*R*,*Z*) poloidal plane, the vertical plane containing the plasma cross-section. $2\pi\psi$ is the flux of the poloidal magnetic field on the plasma mid-plane. The toroidal function *F* corresponds to the amount of poloidal currents inside the plasma (these currents run in the vertical plane). Since they generate a local toroidal magnetic field, *F* also corresponds to the amount of total toroidal field at a radius *R* via the simple relation $F = RB_\phi$. Owing to the non-linear nature of Eq. (5), splitting the plasma equilibrium is not a trivial task. However when the plasma rim is already in equilibrium and its core is a "GSh vacuum" (i.e. no currents, no pressure gradients), then splitting the equilibrium becomes relatively simple. In this particular instance, the innermost flux surface of the rim equilibrium plays the role of a fixed boundary to the GSh equation and an inner equilibrium can be computed in the core "vacuum" using conventional methods. However poloidal field continuity has to be enforced at the interface between both equilibria. Such rim equilibria are called experimentally a current hole [7, 8] as there is no current in the plasma core, only constant pressure

Consequently, to demonstrate the existence of such a composite, or dual, equilibrium, we solved the GSh equation numerically, using the free-boundary code CUBE [ 9 ]. This study will use the following parameters. The plasma major radius *R* is 6 m, the plasma minor radius *a* is 2 m. The plasma elongation factor is 2, triangularity is 0.6 and squareness is 0.1. The toroidal field is 2.5 T at *R* = 6 m. This design resembles ITER's [ 10 ], but uses half its nominal magnetic field. First, we will use a simple model where the current density profile is constant in the plasma rim, reproducing the profile of an artificial current hole equilibrium. While this model is entirely hypothetical, it is a simple template which helps to understand the basic properties of dual equilibria. The inboard (left) part of the core current profile will be set to zero, guarantying a diamagnetic (asymptotic) configuration [5]. Figure 1-a and Figure 1-b present the flux surfaces and toroidal current density distributions across the plasma section. The sharp transition between the rim and core solution is striking and does not jeopardize the existence of an equilibrium. The shape of the current density profile is plotted in the top part of Figure 1-c. Unlike a typical current hole, the absence of toroidal current in the core equilibrium is not synonymous with constant pressure. On the contrary, the inner asymptotic equilibrium has large poloidal currents, which balance pressure gradients in this region. Figure 1-c shows the pressure profile, where gradients are clearly visible inside the core region, and $F$ profile, where both paramagnetism and diamagnetism appear. It is also customary to display some profiles versus the poloidal flux. Since $p$ and $F$ are constant on flux surfaces, the flux space representation is more compact as we plot only the outer half of the $p$ and $F$ functions. In Figure 1-d the flux has been renormalized. As a result, 0 designates the plasma axis (where $p$ is maximum) and 1 the plasma edge. It is also instructive to plot the $q$ profile. This quantity, called the safety factor, represents the ratio of toroidal turns to poloidal turns done by a fictitious collisionless particle traveling on a given flux surface.

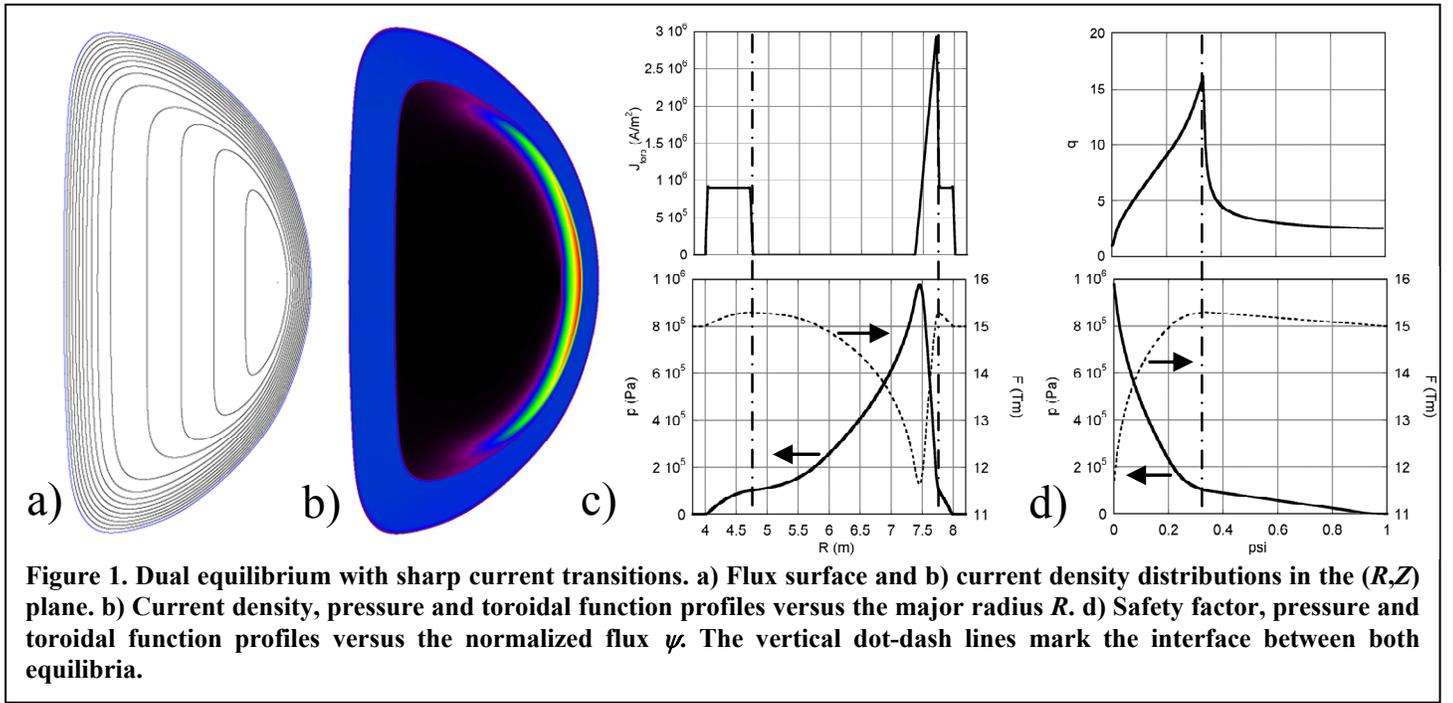

Figure 1. Dual equilibrium with sharp current transitions. a) Flux surface and b) current density distributions in the ($R,Z$) plane. b) Current density, pressure and toroidal function profiles versus the major radius $R$. d) Safety factor, pressure and toroidal function profiles versus the normalized flux $\psi$. The vertical dot-dash lines mark the interface between both equilibria.

The separation between rim and core occurs (evidently) at the sudden current transitions, which coincides with the optimum in $F$ and $q$. The dual nature of the equilibrium appears very clearly on the $q$ profile, which is a good indicator of the location of the core-rim interface.

While this equilibrium perfectly illustrates the "integration by parts" idea previously discussed, its stability will be compromised by the sharp transitions in current densities. As a consequence smoothing is necessary. While many transformations can be envisaged, we decided to retain the most important features of the dual equilibrium, namely toroidal current localization at the plasma edge and inboard-outboard toroidal current asymmetry. Figure 2 shows an instance of possible dual equilibria. This particular current profile was retained since it shares strong similarities with experimental current holes except for the distinctive asymmetry in the current wing heights (current density localization is a given in current hole topologies). Despite strong smoothing, the characteristics found in the previous equilibrium are still present. The two different flux surface distributions (Figure 2-a), the two different $p$ and $F$ profile evolutions (Figure 2-c) with paramagnetic edge and diamagnetic core and the dual $q$ profile (Figure 2-d) are clearly visible. However, the interface between both equilibria is now more difficult to identify due to smoothing. Fortunately the $q$ profile points directly to the interface location, which can be found where the value of $F$ in the plasma corresponds to the value of $F$ in vacuum (namely 15 Tm). In other word, the transition happens where the plasma turns from paramagnetic to diamagnetic ($\psi \sim 0.35$). For this particular equilibrium, the peak $\beta$ (at the location where pressure is maximum) is 100 % and $<\beta>$ is 12 % for a total plasma current $I_P$ of 13 MA. Thus the fusion power computed from Eq. (1) would be similar to ITER ($<\beta> \sim 3\%$, $B_\phi \sim 5$ T), while using only half the magnetic field. The peak pressure is 1 MPa, which is on the order of ITER's. The $\beta_N$ for such an equilibrium is 4.6, a value already obtained experimentally in a finite aspect ratio machine [11].

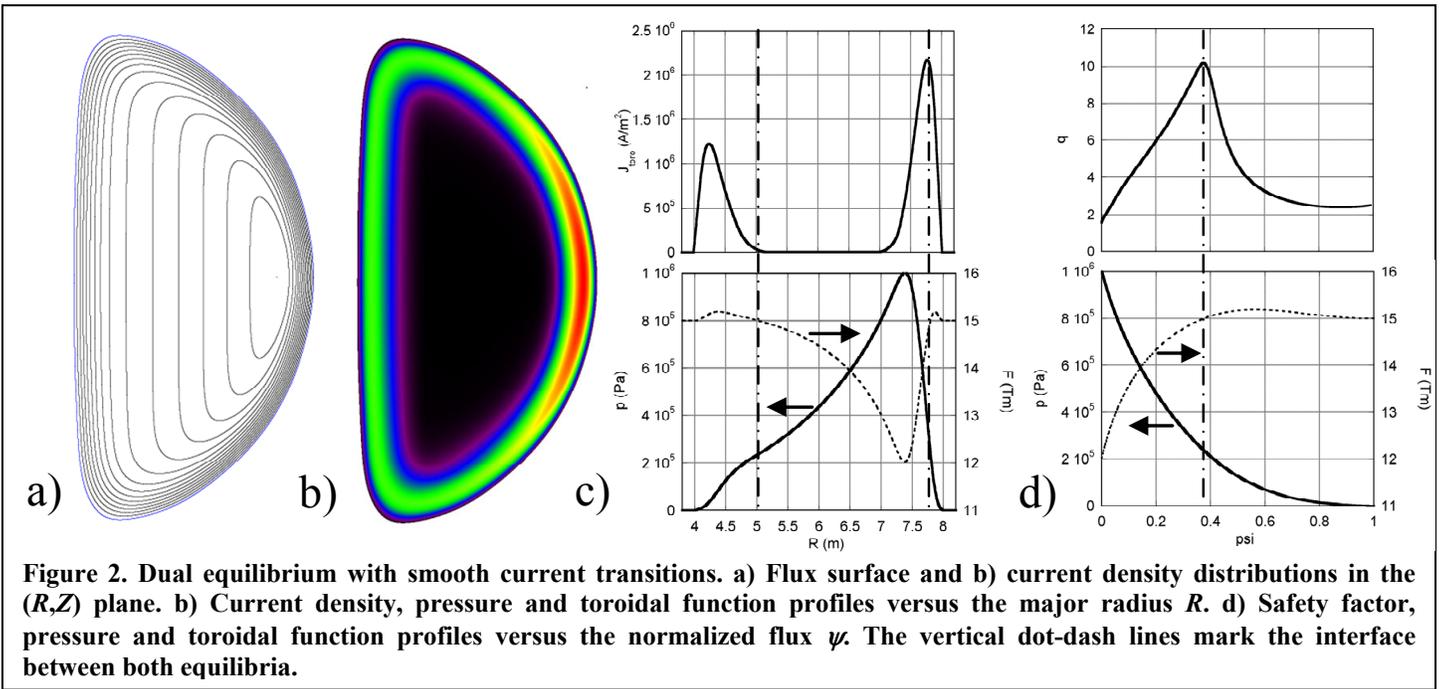

**Figure 2.** Dual equilibrium with smooth current transitions. a) Flux surface and b) current density distributions in the $(R,Z)$ plane. b) Current density, pressure and toroidal function profiles versus the major radius $R$. d) Safety factor, pressure and toroidal function profiles versus the normalized flux $\psi$. The vertical dot-dash lines mark the interface between both equilibria.

This dual equilibrium is more appropriate to stability studies. The stability is investigated numerically with the DCON code [12]. Figure 3-a shows high-$n$ ballooning [13, 14] as well as Mercier [15] stability. Figure 3-b focuses on the stability of the toroidal mode number $n = 1$ for both fixed and free boundary modes. We included in this study all the poloidal harmonics spanning $m = -30$ to $m = 30$. Figure 3-b shows fixed boundary mode stability. While the criterion behavior changes near the interface location, fixed boundary mode stability is present in both equilibria. It is interesting to dwell on the free-boundary mode stability since this is the major issue such high pressure plasmas face. To understand the stabilizing mechanisms, we moved the numerical last closed flux surface of the plasma, supposing vacuum beyond, from the plasma core all the way to the edge. While this approach is not entirely physical, it helps to illustrate how the plasma rim stabilizes free-boundary modes. As we cross the interface between both core and rim equilibria, the change in plasma energy evolution is clearly observable. By extrapolating the core plasma energy all the way to the plasma edge we see that free boundary mode stability is compromised, a behavior already observed in asymptotic equilibria [16]. On the other hand, the presence of the paramagnetic padding changes completely the plasma energy evolution. As the numerical last closed flux surface moves outward, the plasma energy rises rapidly. After we pass the optimum in $F$, located at $\psi = 0.56$, the increase in plasma energy slows down, demonstrating the strong influence of the magnetic field on free-boundary mode stability. The plasma energy at the edge is marginally positive. When the vacuum energy is added to the plasma energy, the total perturbed energy becomes positive for $\psi > 0.75$, guaranteeing free-boundary mode stability for the external kink $n = 1$. This approach demonstrates the influence of the paramagnetic rim and verified the plasma energy dependence with magnetic field. We also found that fixed and free boundary modes for $n = 2$ and 3 are stable in DCON. However, other stability codes showed inconsistencies in stability results. Such discrepancies are expected since high pressure equilibria push stability codes into unexplored

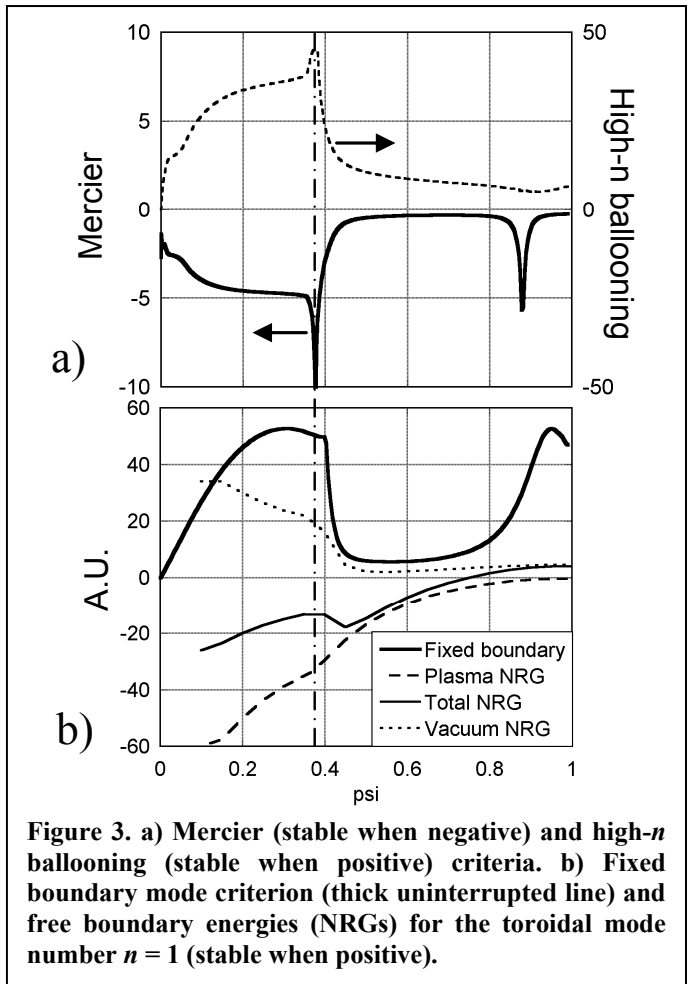

**Figure 3.** a) Mercier (stable when negative) and high-$n$ ballooning (stable when positive) criteria. b) Fixed boundary mode criterion (thick uninterrupted line) and free boundary energies (NRGs) for the toroidal mode number $n = 1$ (stable when positive).

parameter spaces where code artifacts may be detrimental.

In conclusion, this paper presented a new type of unity $\beta$ configuration by combining two equilibria. This dual equilibrium is composed of a diamagnetic core, confining high plasma pressures, and an outer paramagnetic rim, acting as a "perfect conducting wall", stabilizing of the free-boundary modes with toroidal mode numbers $n$ = 1, 2 and 3. Hitherto stability results have to be carefully interpreted. The dual equilibrium has peculiar features such as a dual $q$ profile, which can yield coordinate mapping problems, or large gradients, requiring high resolution of the computational grid. A remarkable property of the dual equilibria is their low $\beta_N$s in elongated configurations. Plasma shaping has radical effects since all the current runs near the edge, where shaping characteristics are most pronounced. Overall, the major asset of dual equilibria is the similarity they share with regular current holes [7, 8]. To our knowledge, an experiment would have to go through a great deal of contortions to obtain unity $\beta$ plasmas when starting from the usual "bell-shaped" current profile [16]. In comparison, reducing the height of the inboard wing of a current hole configuration seems a reasonable strategy that may be easily implemented in present day experiments.

One of the authors (P.-A.G.) would like to thank Dr. Alan Glasser from Los Alamos National Laboratory as well as Drs. Alan Turnbull and Phil Snyder from General Atomics for their help on stability issues.